\documentclass[conference]{IEEEtran}
\usepackage{epsfig}
\usepackage{times}
\usepackage{float}
\usepackage{afterpage}
\usepackage{amsmath}
\usepackage{amstext}
\usepackage{amssymb,bm}
\usepackage{latexsym}
\usepackage{color}
\usepackage{graphicx}
\usepackage{amsmath}
\usepackage{amsthm}
\usepackage{graphicx}
\usepackage[center]{caption}
\usepackage{pstricks}
\usepackage{caption}
\usepackage{subcaption}
\usepackage{booktabs}
\usepackage{multicol}
\usepackage{lipsum}
 \usepackage[T1]{fontenc}
\usepackage{epstopdf}

\allowdisplaybreaks

\theoremstyle{definition}

\providecommand{\definitionname}{Definition}

\newfloat{algorithm}{tbp}{loa}
\providecommand{\algorithmname}{Algorithm}
\floatname{algorithm}{\protect\algorithmname}

\begin{document}

\title{Novel Delivery Schemes for Decentralized Coded Caching in the Finite File Size Regime}

\author{\IEEEauthorblockN{Kai~Wan}\IEEEauthorblockA{Laboratoire des Signaux et Système (L2S)\\
CentraleSupélec-CNRS-Université Paris-Sud\\
Gif-sur-Yvette, France\\
Email: kai.wan@u-psud.fr}\and \IEEEauthorblockN{Daniela~Tuninetti}\IEEEauthorblockA{NICEST Lab\\University of Illinois at Chicago\\
Chicago, IL 60607, USA\\
Email: danielat@uic.edu}\and \IEEEauthorblockN{Pablo~Piantanida}\IEEEauthorblockA{Laboratoire des Signaux et Système (L2S)\\
CentraleSupélec-CNRS-Université Paris-Sud\\
Gif-sur-Yvette, France\\
Email: pablo.piantanida@centralesupelec.fr}}

\maketitle

\begin{abstract}
This paper analyzes the achievable tradeoff between cache~size and download~rate in decentralized caching systems with the uncoded cache placement originally proposed by Maddah-Ali and Niesen. It proposes two novel delivery schemes that take advantage of the multicasting opportunities that arise when a file is demanded by multiple users. These delivery schemes are extensions of known ones to the regime where the file size is finite. Numerical evaluations for the case of file uniform popularity show that the proposed schemes outperform previous ones for all value of the cache size.
\end{abstract}

\IEEEpeerreviewmaketitle{}

\section{Introduction}
\label{sec:intro}

The fundamental limits of cache-aided systems were studied by Maddah-Ali and Niesen (MAN) in~\cite{dvbt2fundamental,decentralizedcoded}. In the MAN model, a server is connected to $K$~users, or clients, via a shared error-free broadcast link. The server has a library of $N$~files, each of size $F$~bits. Each user has a local cache of size $MF$~bits to store parts of the files available in the library. There are two phases in a cache system. In the {\it placement phase}, pieces of the files are stored within the users' cache 
without knowledge of the future user demands. 
In the {\it delivery phase}, 
the server, based on the users' demands and the cache contents, 
broadcasts packets 
to all users so that each user can recover the demanded file.  
The objective 
is to design a two-phase scheme 
that minimizes the number of transmitted packets in the delivery phase. 
If the $K$ connected users are not the same during both phases, e.g., due to the user's mobility, each server must carry out independently the placement phase. 
In this paper, for practical reasons, we 
consider only decentralized caching systems.

\paragraph*{Past Work for $F\to+\infty$}
In~\cite{decentralizedcoded} Maddah-Ali and Niesen proposed a decentralized caching scheme, which we refer to as {\it decMAN}, where each user fills its cache randomly and independently of the others.
After the placement phase, the bits of each file can be divided into sub-files depending on the users who have them in their cache.  
In delivery phase, 
each sub-file 
is treated as a district message and delivered by using a linear code as in the centralized MAN scheme~\cite{dvbt2fundamental}. 
The exact rate-memory tradeoffs for decMAN 
with {\it uniform demands} was given in~\cite{exactrateuncoded},  
where the authors proved that some linear combinations in the original decMAN are reduandant.

For non-uniform demands, the authors of~\cite{nonuniformdemands} proposed to divide the files into groups depending on their popularity, where the files in the same group taking the same amount of cache space, and deliver them as in the original decMAN.
In~\cite{detailsgroupe}, the authors proposed a new (compared to decMAN) placement phase 
and a delivery phase based on graph colouring. 
With the placement phase as in~\cite{detailsgroupe}, the authors of~\cite{simplifiedgroupeindexcoding} proposed an approximate method to simply the computation of the local chromatic number.
In~\cite{deliveryscheme} a delivery scheme based on decMAN, but independent of placement policy, was proposed.
With the caching policies for non-uniform demand, the sub-files
in a linear combination of {decMAN} may have different sizes;
instead of zero-padding 
the shorter sub-files, the scheme in~\cite{deliveryscheme} pads with bits from other sub-files that need to be transmitted. 
%

In this paper we consider the more practical case of finite file size. Note that by using the above schemes directly in the finite file size regime, except the scheme in~\cite{deliveryscheme}, those schemes should be modified. For the schemes in~\cite{decentralizedcoded,exactrateuncoded,nonuniformdemands}, since  when $F$ is finite the sub-files in each linear combination may have different sizes, zero-padding should be used on the shorter ones to meet the length of the longest one. 

\paragraph*{Past Work for finite $F$}
The finite file size regime was considered in~\cite{finiteanalysis}, where a slightly modified caching scheme (compared to the original decMAN) was shown to get a multiplicative gain when the file size $F$ grows exponentially with this gain. In~\cite{newfinitefilesize}, the authors proposed a caching scheme that outperforms decMAN when $F$ is not large.

\paragraph*{Our Contribution}
In this paper, we investigate the memory-load tradeoff for decentralized caching systems with finite file size. The schemes in~\cite{deliveryscheme,finiteanalysis,newfinitefilesize} treat each bit demanded by each user as a district one thus
not leveraging the multicasting opportunities that arise when the same bits are demanded by several users. 
We propose two delivery schemes, independent on the used placement phase, based on the schemes in~\cite{simplifiedgroupeindexcoding} and~\cite{deliveryscheme}, respectively. The first proposed scheme, which can be seen as an advanced version of the delivery algorithm in~\cite{simplifiedgroupeindexcoding}, while the second proposed scheme shares with~\cite{deliveryscheme} the main idea in order to adapt the scheme in~\cite{exactrateuncoded} to the finite file size regime.
Numerical evaluations for the case of 
uniform demands show that our proposed schemes outperform existing ones
for finite file size regime and every value of the cache memory.


\paragraph*{Paper Outline}
The rest of the paper is organized as follows. 
Section~\ref{sec:model} presents the system model.
Section~\ref{sec:newach} presents the two proposed delivery schemes.
Section~\ref{sec:num} 
presents numerical results and complexity considerations.
Finally, Section~\ref{sec:conclusions} concludes the paper.

\paragraph*{Notation}
Calligraphic symbols denote sets,
$|\cdot|$ the cardinality of a set or the length of a file,
$[n_1:n_2]$ the set of integers from $n_1$ to $n_2$, and
$\oplus$  the bit-wise XOR operation.

\section{Problem Formulation and an Achievable Load} 
\label{sec:model}

Firstly we  define the decentralized caching problem with finite file size $F$.
Secondly we revise the scheme in~\cite{exactrateuncoded}, which is optimal for uniform demands when $F\to\infty$ and will be used as an outer bound in the numerical evaluations.

\subsection{Problem Statement}
\label{subsection:Problem Statement}
We consider a decentralized 
caching system where 
a server with $N$ files, denoted by $(F_{1},F_{2},\dots,F_{N})$, is connected to $K$ users through a shared error-free link.  Each file has $F$ bits. 
In the placement phase, user $i\in[1:K]$ stores some bits from the $N$~files in his cache of size $MF$~bits without knowledge of later demands, where $M\in[0,N]$. We denote the content of the cache of user $i$ as $Z_{i}$, and let $\mathbf{Z}:=(Z_{1},\dots,Z_{K})$. 
In decentralized systems 
where coordination among users' placements is not allowed, the caching functions are arbitrary;
based on one caching strategy denoted by $\mathcal{C}(M)$, the realization of cache configuration $\mathbf{Z}$ is also arbitrary.
In the delivery phase, each user demands one file and the demand vector $\mathbf{d}:=(d_{1},d_{2},\dots,d_{K})$ is revealed to all parties, where $d_{k}\in[1:N]$ corresponds to the file demanded by user $k\in[1:K]$. 
Let $\mathcal{N}(\mathbf{d})$ be the set of distinct files in the demand vector $\mathbf{d}$.
Given $(\mathbf{d},\mathbf{Z})$, the server broadcasts a message $X_{\mathbf{d},\mathbf{Z}}$ of length 
$R(\mathbf{d},\mathbf{Z}) F$ bits.
With $X_{\mathbf{d},\mathbf{Z}}$ and $Z_{k}$ each user $k\in [1:K]$ recovers his demanded file without error.
The objective is to minimize the {\it average network load}
\begin{equation}
R^{*}(M,F)
:=\min_{\mathcal{C}(M)}\mathbb{E}_{\mathbf{d},\mathbf{Z}}[R(\mathbf{d},\mathbf{Z})],
\label{eq:R*(M)}
\end{equation}
where the expectation is over the distribution induced on $\mathbf{Z}$ by the adopted caching strategy $\mathcal{C}(M)$ and 
the distribution on the demand vector $\mathbf{d}$.

\subsection{Case of uniform demands in the asymptotic regime}
\label{subsection:A Deliver Scheme for F=infty}

In the placement phase of decMAN, each user 
randomly and uniformly stores $\frac{MF}{N}$ bits of file $F_{i}$ for $i\in[1:N]$.

Given the cache content of all the users, the bits of the files are grouped into sub-files $F_{i,\mathcal{W}}$,
where $F_{i,\mathcal{W}}$ is the set of bits of file $i\in[1:N]$ known by the users in $\mathcal{W}\subseteq[1:K]$. 
Since $F\to\infty$, by the Law of Large Numbers, the length of each sub-file only depends on the number of users who know it, that is, for all $i\in[1:N]$ we have
\begin{align}
\frac{|F_{i,\mathcal{W}}|}{F}\to & \left(\frac{M}{N}\right)^{|\mathcal{W}|}\left(1-\frac{M}{N}\right)^{K-|\mathcal{W}|},\,\,\textrm{in probability}. 
\label{eq:law of large number}
\end{align}

In the original decMAN, the server transmits $X_{\mathcal{S}} := \oplus_{s\in\mathcal{S}}F_{d_{s},\mathcal{S}\backslash\{s\}}$ for all $\mathcal{S}\subseteq[1:K]$ where $\mathcal{S}\neq \emptyset$, so that 
user $i\in \mathcal{S}$ can recover $F_{d_{i},\mathcal{S}\backslash\{i\}}$ from $X_{\mathcal{S}}$. 
%
In~\cite{exactrateuncoded} it was shown that for each $t\in[0:K-1]$, among all $\binom{K}{t+1}$ coded messages $X_{\mathcal{S}}$
with 
$|\mathcal{S}|=t+1$,
$\binom{K-|\mathcal{N}(\mathbf{d})|}{t+1}$ of them can be obtained by linear combinations of the remaining ones and thus need not be transmitted. 
%
In particular, for each file $i\in\mathcal{N}(\mathbf{d})$, 
we randomly choose a `leader user' demanding this file and add this user to the set $\mathcal{L}$; 
then we transmit all $X_{\mathcal{S}}$ with
$\mathcal{S}\cap\mathcal{L}\neq\emptyset$. 
The average load is thus bounded by~\cite[Theorem~2]{exactrateuncoded}
\begin{align}
R^{*}(M)\leq\frac{N-M}{M}\mathbb{E}\left[1-\left(\frac{N-M}{N}\right)^{|\mathcal{N}(\mathbf{d})|}\right].
\label{eq:exactrateuncoded result}
\end{align}
In~\cite{exactrateuncoded} a matching converse  to~\eqref{eq:exactrateuncoded result} was provided for the case of uniform demands.

We conclude this section by noting that the main difference between decMAN and the scheme in~\cite{exactrateuncoded} is that in the delivery phase decMAN treats each sub-file demanded by each user as a distinct sub-file; thus decMAN does not profit from the multicasting opportunities that arise when sub-files are demanded by several users, which instead is what~\cite{exactrateuncoded} does.


\section{Two Novel Delivery Schemes}
\label{sec:newach}
In this section we introduce two delivery algorithms for the finite file size regime that
work for {\it any} placement phase and {\it any} demand distribution.
In principle, {decMAN} and the scheme in~\cite{exactrateuncoded} for $F\to+\infty$ can be applied to the case of finite $F$ as follows: if the sub-files involved in a linear combination do not have the same length, it suffices to zero~pad the shorter sub-files to match the length of the longest one.
This simple trick may results in unnecessary transmissions. 
In addition, since one file may be demanded by several users, there are multicasting opportunities 
to leverage.
Based on these two ideas, 
we extend the Hierarchical greedy Local Colouring algorithm (HgLC) of~\cite{simplifiedgroupeindexcoding} and the Heterogenous Coded Delivery (HCD) of~\cite{deliveryscheme},  
which weres originally proposed for the infinite file size regime. 
%


\subsection{Delivery Scheme~1: AHgLC}
\label{sub:HgLC}

The placement proposed in~\cite{simplifiedgroupeindexcoding} divides each file into $B$ packets of length $F/B$. Each user randomly and uniformly chooses $MB/N$ packets of each file to be stored. 
After the placement phase, the authors of~\cite{simplifiedgroupeindexcoding} proposed to generate a directed graph $\mathcal{H}$ as follows. Define $\mathcal{V}$ as the node set and $\mathcal{E}$ as the edge set.
Each packet requested by a user who does not cache it is a distinct vertex $v$ in the graph.
The user requesting $v$ is denoted by $\mu(v)$ while the packet corresponding to $v$ is denoted by $\rho(v)$. 
Note that different vertices may correspond to the same packet.
A direct edge from $v_{2}$ to $v_{1}$ exists if  $\mu(v_{2})$ does not cache $\rho(v_{1})$, and $\rho(v_{1})$ and $\rho(v_2)$ do not represent the same packet. The number of transmissions needed to satisfy all users is equal to the {\it local chromatic number} of this directed graph, which is NP-hard to compute.
%
In~\cite{simplifiedgroupeindexcoding}, an approximate algorithm, referred to as Hierarchical greedy Local Colouring algorithm (HgLC), was proposed in order to simply numerical computations.
HgLC works as follows. 
For each node $v$ let $\mathcal{K}_{v}$ 
be the set of the users who either demand $\rho(v)$ 
or have $\rho(v)$ in the cache.
Divide the nodes into hierarchies, where the $i$-th~hierarchy is $\mathcal{G}_{i}=\{v:|\mathcal{K}_{v}|=i\}, \ i\in [1:K]$. 
Then, run a loop from the highest 
to the lowest hierarchy; 
at the step for the $i$-th~hierarchy, for each node $v\in \mathcal{G}_{i}$ 
find the largest sets of non-adjacent nodes in $\mathcal{G}_{i}$ containing $v$;
if the length of the found set is not less than $i$, then color all the nodes in this set by a new color and remove these nodes from $\mathcal{G}_{i}$; otherwise, remove $v$ from $\mathcal{G}_{i}$ and add it into $\mathcal{G}_{i-1}$. After colouring all the nodes in the graph, since $F/B$ can be arbitrary large enough in the infinite file size regime, an MDS 
code is used for the local colouring.

The pseudo-code for the proposed extension of HgLC, referred to as Advanced Hierarchy greedy Local Colouring Algorithm (AHgLC), is given in Algorithm 1. 
The first improvement compared to HgLC is a novel way for searching unconnected sets. In contrast to HgLC, we do the iteration from the lowest hierarchy $1$ to the highest hierarchy $K$. In addition, for each node $v$, instead of randomly searching unconnected nodes, we firstly search the nodes $w_{1}$ where $\mathcal{K}_{w_{1}}\supseteq\mathcal{K}_{v}$ then the nodes $w_{2}$ where $\mathcal{K}_{w_{2}}\nsupseteq\mathcal{K}_{v}$. The second improvement is to do the local colouring by random linear binary combinations, in contrast to HgLC which uses the parity-check matrix of MDS code with a large field size when the code length is large. In the following, we introduce the two improvements in details.

{\it Improvement~1.} For one node $v$,  there are $|\mathcal{K}_{v}|$ users either knowing the packet $\rho(v)$ or
demanding it. Since the nodes demanded by the same user are surely connected, the length of the largest possible unconnected set containing
$v$ is $|\mathcal{K}_{v}|$. In other words, while transmitting $v$, we can transmit
at most $|\mathcal{K}_{v}|-1$ other nodes at the same time (i.e.,
with one color). In addition, for each node $w$ in an unconnected set containing $v$ with length $|\mathcal{K}_{v}|$, we have $\mathcal{K}_{w}\supseteq\mathcal{K}_{v}$. Hence, we can transmit some nodes in the same or higher hierarchies than $v$ while transmitting the largest unconnected set containing $v$. So if we do the iteration from the highest hierarchy to the lowest hierarchy as HgLC does, before the low level nodes we may transmit 
high hierarchy nodes which can be transmitted with the low level nodes. 

For example, consider the case with $K=N=4$ and the demand vector $\mathbf{d}=\{1,2,3,4\}$;
there are $7$ sub-files with equal length 
$F_{d_{1},\{2,3\}}$, $F_{d_{2},\{1,3\}}$, $F_{d_{3},\{1,2\}}$, $F_{d_{1},\{3\}}$, $F_{d_{2},\{3\}}$, $F_{d_{3},\{2\}}$, $F_{d_{4},\emptyset}$;
the iteration from high to low hierarchies gives the code $F_{d_{1},\{2,3\}}\oplus F_{d_{2},\{1,3\}} \oplus F_{d_{3},\{1,2\}}$, $F_{d_{1},\{3\}}$, $F_{d_{2},\{3\}}$, $F_{d_{3},\{2\}}$, $F_{d_{4},\emptyset}$ of length $5$,
while the inverse order gives $F_{d_{1},\{3\}}\oplus F_{d_{3},\{1,2\}}$, $F_{d_{2},\{3\}}\oplus F_{d_{1},\{2,3\}}$, $F_{d_{3},\{2\}}\oplus F_{d_{2},\{1,3\}}$, $F_{d_{4},\emptyset}$ of length $4$.

For each node in one hierarchy we want to find the unconnected set $\mathcal{I}$ with length $|\mathcal{K}_{v}|-m$ containing it, where $m$ is initially  $0$ meaning that we desire to find the largest set. After one loop, we increase $m$ by $1$ to relax the length constraint of the sought unconnected set and colour the uncoloured nodes. For node $v$,  we firstly add the nodes representing the same packet to $\mathcal{I}$ (Step (2.c.i) in Algorithm 1), then do the search among all the nodes $w$ in the same hierarchy where $\mathcal{K}_{w}\supseteq\mathcal{K}_{v}$ (Step (2.c.ii) to (2.c.iv) in Algorithm 1). If the found set does not reach the length constraint, we search the nodes $w$ in the next hierarchy where $\mathcal{K}_{w}\supseteq\mathcal{K}_{v}$. If after searching all the nodes $w$ in all the hierarchies   where $\mathcal{K}_{w}\supseteq\mathcal{K}_{v}$, the found set is still smaller than the constraint, we search the nodes in $\{w:\mathcal{K}_{w}\nsupseteq\mathcal{K}_{v}\}$ (Step (2.e) in Algorithm 1). If the final found set reach the length constraint, we colour this set by a new colour and take its nodes out of the hierarchies. If not, we search the unconnected set for the remaining nodes.

{\it Improvement~2.} Denote the colouring of this graph $\mathcal{H}$ by $\mathbf{c}_{\mathcal{H}}$ and the number of used colors by $|\mathbf{c}_{\mathcal{H}}|$. Denote the packets corresponding to color $c$ by $\mathcal{V}_{c}=\{\rho(v):\textrm{ node }v\textrm{ is colored by }c\}$. For each color $c$,  we transmit $\oplus_{p\in\mathcal{V}_{c}}p$. By doing so, the code $\mathbf{G}\times\mathbf{P}$ can be generated for this graph colouring, where $\mathbf{P}$ of dimension $NF\times 1$ represents all the bits to transmit and the dimension of $\mathbf{G}$ is $|\mathbf{c}_{\mathcal{H}}|\times NF$. We then do the local colouring by random linear combinations.  
For each $i\in[1:K]$, $\mathcal{A}_{i}$ represents the set of all the colors $c$  where all the packets in $\mathcal{V}_{c}$ are known by user $i$. Hence, we should construct a binary matrix $\mathbf{C}$ of dimension $(|\mathbf{c}_{\mathcal{H}}|-\min_{i\in[1:K]}|\mathcal{A}_{i}|)\times|\mathbf{c}_{\mathcal{H}}|$ such that for each user $i\in[1:K]$, the matrix formed by  
columns indexed by the complement of $\mathcal{A}_{i}$ of $\mathbf{C}$ has a rank $|\mathbf{c}_{\mathcal{H}}|-|\mathcal{A}_{i}|$. If we can construct such matrix, each user can recover $\mathbf{G}\times\mathbf{P}$ from $\mathbf{C}\times \mathbf{G}\times\mathbf{P}$ and so we transmit $\mathbf{C}\times \mathbf{G}\times\mathbf{P}$. Otherwise, we transmit $\mathbf{G}\times\mathbf{P}$. Algorithm 2 is used to construct such binary matrix. 
Numerically we noted that such matrix can be always constructed  if $|\mathbf{c}_{\mathcal{H}}|-\min_{i\in[1:K]}|\mathcal{A}_{i}|>K$.

\rule{0.45\textwidth}{0.8pt}

\vspace{-3bp}
\textbf{Algorithm 1 }AHgLC
\vspace{-6bp}

\rule{0.45\textwidth}{0.8pt}
\begin{enumerate}
\item \textbf{Input}: $\mathcal{G}_{i}$ for all $i\in[1:K]$, the color
set $\mathbf{c}_{\mathcal{H}}=\emptyset$, $m=0$;
\item \textbf{for} $i=1,\ldots,T$ where $\mathcal{G}_{i}\neq\emptyset$,
\begin{enumerate}
\item $\mathcal{Q}=\mathcal{G}_{i}$;
\item randomly pick a node $v$ from $\mathcal{Q}$; $\mathcal{I}=\{v\}$;
$Q=Q\setminus\{v\}$;
\item \textbf{for} $j=i,i+1,\ldots,T$ where $\mathcal{G}_{j}\neq\emptyset$,
\begin{enumerate}
\item $\mathcal{I}=\mathcal{I}\cup\{u\in\mathcal{G}_{j}:\rho(v)=\rho(u)\};$.
\item $\mathcal{W}=\{w\in\mathcal{G}_{j}\setminus\mathcal{I}:\mathcal{K}_{w}\supseteq\mathcal{K}_{v}\};$
\item randomly pick a node $w$ from $\mathcal{W}$; $\mathcal{W}=\mathcal{W}\setminus\{w\}$;
\item \textbf{if} there is no edge between $w$ and $\mathcal{I}$, \textbf{then} $\mathcal{I}=\mathcal{I}\cup\{w\}$;
\item \textbf{if} $\mathcal{W}\neq \emptyset$, \textbf{then} go to Step (2.c.iii);
\end{enumerate}
\item \textbf{if} $\mathcal{I}\geq i-m$, go to Step (2.f); 
\item \textbf{for} $j=i,i+1,\ldots,T$ where $\mathcal{G}_{j}\neq\emptyset$,
\begin{enumerate}
\item $\mathcal{W}=\{w\in\mathcal{G}_{j}\setminus\mathcal{I}:\mathcal{K}_{w}\nsupseteq\mathcal{K}_{v}\};$
\item randomly pick a node $w$ from $\mathcal{W}$; $\mathcal{W}=\mathcal{W}\setminus\{w\}$;
\item \textbf{if} there is no edge between $w$ and $\mathcal{I}$, \textbf{then} $\mathcal{I}=\mathcal{I}\cup\{w\}$;
\item \textbf{if} $\mathcal{W}\neq \emptyset$, \textbf{then} go to Step (2.e.ii); 
\end{enumerate}
\item \textbf{if} $\mathcal{I}\geq i-m$, \textbf{then}
\begin{enumerate}
\item color all the vertices in $\mathcal{I}$ by $|\mathbf{c}_{\mathcal{H}}|+1;$
\item $\mathbf{c}_{\mathcal{H}}=\mathbf{c}_{\mathcal{H}}\cup\{|\mathbf{c}_{\mathcal{H}}|+1\};$ $\mathcal{Q}=\mathcal{Q}\setminus\mathcal{I};$
\item \textbf{for} each $n\in\mathcal{I}$,
\begin{enumerate}
\item $\mathcal{G}_{|\mathcal{K}_{n}|}=\mathcal{G}_{|\mathcal{K}_{n}|}\setminus \{n\};$
\end{enumerate}
\end{enumerate}
\item \textbf{if} $\mathcal{Q}\neq \emptyset$, \textbf{then} go to Step (2.b);
\end{enumerate}
\item \textbf{if} $\exists\thinspace\thinspace\textrm{one}\thinspace\thinspace i\in[1:T],\thinspace\thinspace\textrm{s.t}.\thinspace\thinspace\mathcal{G}_{i}\neq \emptyset$, \textbf{then} 
\begin{enumerate}
\item $m=m+1$;
\item go to Step (2);
\end{enumerate}
\item \textbf{for} each $c\in \mathbf{c}_{\mathcal{H}}$,
\begin{enumerate}
\item $\mathcal{V}_{c}=\{\rho(v):\textrm{ node }v\textrm{ is colored by }c\}$;
\item $\textrm{the code corresponds to this color is }\oplus_{p\in\mathcal{V}_{c}}p$;
\end{enumerate}
\item denote the code corresponding to the above colouring by $\mathbf{G}$;
\item \textbf{for} each $i\in[1:K]$,
\begin{enumerate}
\item $\mathcal{A}_{i}=\{c\in\mathbf{c}_{\mathcal{H}}:\mathcal{V}_{c}\textrm{ is known by user }i\}$;
\end{enumerate}
\item {Output} $RLC(|\mathbf{c}_{\mathcal{H}}|,\mathcal{A}_{1},\ldots,\mathcal{A}_{K})\times\mathbf{G}\times \mathbf{P}$;
\end{enumerate}

\rule{0.45\textwidth}{0.8pt}
 
\rule{0.45\textwidth}{0.8pt}

\vspace{-3bp}
\textbf{Algorithm 2} $RLC(L,,\mathcal{A}_{1},\ldots,\mathcal{A}_{K})$
\vspace{-6bp}

\rule{0.45\textwidth}{0.8pt}
\begin{enumerate}
\item {Input}: $L$, $\mathcal{A}_{1},\ldots,\mathcal{A}_{K}$. \textbf{Initialization}: $t_{1}=0;$ $times=10;$
\item $Test=1;$ $t_{1}=t_{1}+1$; randomly generate a $(L-\min_{i\in[1:K]}|\mathcal{A}_{i}|)\times L$ binary matrix $\mathbf{C}$;
\item \textbf{for} each $i\in [1:K]$,
\begin{enumerate}
\item form $\mathbf{C}_{1}$ by all the $i^{\textrm{th}}$ columns where $i\in[1:L]\setminus\mathcal{A}_{i}$;
\item \textbf{if} the rank of $\mathbf{C}_{1}$ is less than $m-|\mathcal{A}_{k}|$, \textbf{then}
\begin{enumerate}
\item $Test=0$;
\item \textbf{break for};
\end{enumerate}
\end{enumerate}
\item \textbf{if} $Test=0$ and $t_{1}\leq times$, \textbf{then} go to Step (2);
\item \textbf{if} $Test=1$, \textbf{then} {Output} $\mathbf{C}$; \\
\textbf{else, then }{Output} $\mathbf{I}_{L\times L}$;
\end{enumerate}

\rule{0.45\textwidth}{0.8pt}

\subsection{Delivery Scheme~2: MHCD}
\label{sub:HCD}

When demands are not uniform, the placement depends on the file popularity and the number of stored bits of each file in one user's cache may not be identical. Hence, the sub-files in each $X_{\mathcal{S}}$ where $\mathcal{S}\subseteq[1:K]$ and $\mathcal{S}\neq \emptyset$ defined in Section~\ref{subsection:A Deliver Scheme for F=infty} may have different sizes. Instead of padding zeros at the end of shorter sub-files,  
~\cite{deliveryscheme} proposed a scheme called Heterogenous Coded Delivery (HCD). As decMAN, HCD treats each sub-file demanded by each user as a district one. For each $t\in [0:K-1]$, each $\mathcal{S}\subseteq[1:K]$ of size $|\mathcal{S}|=t+1$ and each $s\in \mathcal{S}$, if $|F_{d_{s},\mathcal{S}\backslash\{s\}}|<\max_{s\in\mathcal{S}}|F_{d_{s},\mathcal{S}\setminus\{s\}}|$, HCD borrows up to $\max_{s\in\mathcal{S}}|F_{d_{s},\mathcal{S}\setminus\{s\}}|-|F_{d_{s},\mathcal{S}\backslash\{s\}}|$ bits from the sub-files $F_{d_{s},\mathcal{J}}$ where $s\notin\mathcal{J}$ and $\mathcal{J}\supseteq\mathcal{S}\setminus\{s\}$. One can see that $F_{d_{s},\mathcal{J}}$ should be recovered by $s$ while it is known by the users in $\mathcal{S}$ except $s$. Hence, the borrowed bits need not to be sent to user $s$ in the later transmission. 

{\it Improvement.} It can be seen that HCD is tailored for the finite file size regime. However, the main limitation of HCD is that it does not leverage the multicasting opportunities arising  from one file demanded by several users.
In the following, we propose a delivery scheme based on~\cite{exactrateuncoded} and HCD. The main difference between our proposed scheme, referred to as (Multicasting Heterogenous Coded Delivery) MHCD,  and HCD is that MHCD adapts the borrowing bits from the higher type sub-files idea to the scheme in~\cite{exactrateuncoded} while HCD adapts it to decMAN. To leverage~\cite{exactrateuncoded}, each sub-file demanded by each user cannot be treated as a district one as decMAN does.  Instead, after the bit borrowing step, each sub-file appearing in different linear combinations should be identical. The following example shows this point. 

Assume that user~$1$ and~$2$ demand file $A$, user~$3$ and~$4$ demand file $B$, and user~$5$,~$6$ and~$7$ demand file $C$. Assume that decMAN and the scheme in~\cite{exactrateuncoded} need to transmit $A_{\{3,4\}}\oplus B_{\{1,4\}}\oplus B_{\{1,3\}}$ and $A_{\{3,4\}}\oplus B_{\{2,4\}}\oplus B_{\{2,3\}}$; 
assume that $A_{\{3,4\}}$ has $1$ bit, 
both $B_{\{1,4\}}$ and $B_{\{1,3\}}$ have $2$ bits, and 
$B_{\{2,4\}}$ and $B_{\{2,3\}}$ have $3$ bits. 
HCD borrows one bit from $A_{\mathcal{J}_{1}}$ where $\mathcal{J}_{1}\supseteq\{3,4\}$ and $1\notin\mathcal{J}_{1}$, and pads this bit at the end of $A_{\{3,4\}}$ in the first linear combination. This borrowed bit need not be sent to user~$1$ in the following transmission. However, if this borrowed bit is also demanded by user~$2$, we should send it to user~$2$ in a later transmission/linear combination.  Similarly, HCD borrows two bits from $A_{\mathcal{J}_{2}}$  where $\mathcal{J}_{2}\supseteq\{3,4\}$ and $2\notin\mathcal{J}_{1}$ and pads these two bits at the end of $A_{\{3,4\}}$ in the second sum. The two borrowed bits need not be sent to user~$2$ in following transmissions. However, if these borrowed  bits are also demanded by user~$1$, we should send them to user~$1$ in some linear combinations later. 
Thus we adapt the borrowing bits idea to the scheme in~\cite{exactrateuncoded}. Since $A_{\{3,4\}}$ in the two sums should be identical, we must pad the same bits from $A_{\mathcal{J}}$ where $\mathcal{J}\supseteq\{3,4\}$ and $\{1,2\}\cap\mathcal{J}_{1}=\emptyset$  at the end of the $A_{\{3,4\}}$ of both sums. The borrowed bits need not to be sent to user~$1$ and~$2$ in the following transmission.

Our proposed MHCD algorithm works as follows. Transmit all the sub-files $F_{i,\emptyset}$ where $i\in\mathcal{N}(\mathbf{d})$. In the following, we consider the sub-files step by step from the ones known by a single user to the ones known by $K-1$ users. MHCD includes $K-1$ steps, from step $1$ to $K-1$. In the remaining of this subsection, we will introduce the procedure of MHCD of step $t$, where all the sub-files $\{F_{d_{i},\mathcal{J}}:i\in[1:K],\mathcal{J}\subseteq[1:K],|\mathcal{J}|=t,i\notin\mathcal{J}\}$ should be transmitted. The pseudo code of MHCD in step $t$ is given in Algorithm~3. 
We need to decide how many bits should be borrowed for each considered sub-file. The key point is that in each linear combination, there exists at least one sub-file without the bits from the higher type sub-files.
Recall $\mathcal{L}$ is the leader set defined in Section~\ref{subsection:A Deliver Scheme for F=infty}. The scheme in~\cite{exactrateuncoded} transmits $\oplus_{s\in\mathcal{S}}F_{d_{s},\mathcal{S}\backslash\{s\}}$ where $\mathcal{S}\in\mathcal{C}_{\mathcal{L}}(t)$ and $\mathcal{C}_{\mathcal{L}}(t)=\{\mathcal{S}\subseteq[1:K]:|\mathcal{S}|=t+1,\mathcal{S}\cap\mathcal{L}\neq\emptyset\}$. We divide all the elements in $\mathcal{C}_{\mathcal{L}}(t)$ into groups,
\begin{equation*}
\mathcal{Q}_{t,\mathcal{T}}=\left\{\mathcal{S}\subseteq\mathcal{C}_{\mathcal{L}}(t):\underset{s\in\mathcal{S}}{\bigcup}\{d_{s}\}=\mathcal{T}\right\}, 
\end{equation*} 
where $\mathcal{T}\subseteq\mathcal{N}(\mathbf{d})$ and $1\leq|\mathcal{T}|\leq\min\{t+1,|\mathcal{N}(\mathbf{d})|\}$. For two different $\mathcal{T}_{1}$ and $\mathcal{T}_{2}$,  it can be seen that $\{(d_{s},\mathcal{\mathcal{S}}\setminus\{s\}):s\in\mathcal{S},\mathcal{S}\in\mathcal{Q}_{t,\mathcal{T}_{1}}\}\cap\{(d_{s},\mathcal{\mathcal{S}}\setminus\{s\}):s\in\mathcal{S},\mathcal{S}\in\mathcal{Q}_{t,\mathcal{T}_{2}}\}=\emptyset$.
Hence, we can encode each group individually such that one sub-file in different sums is identical. In addition, for each $k\in[1:K]$ and each $\mathcal{J}\subseteq[1:K]$ where $k\notin\mathcal{J}$, we define $\mathcal{D}_{k,\mathcal{J}}$ as the {\it individual requested} sub-file representing the set of bits demanded by user $k$ which are not already transmitted and are in the caches of the set of users in $\mathcal{J}$.
Then, define the {\it common requested} sub-file
$$
\mathcal{W}_{i,\mathcal{J}}=\underset{k\in[1:K]:d_{k}=i,k\notin\mathcal{J}}{\bigcap}\mathcal{D}_{k,\mathcal{J}}
$$
representing the common demanded and not yet transmitted bits of all the users who desire $F_{i,\mathcal{J}}$. Initially, $\mathcal{D}_{k,\mathcal{J}}=\mathcal{W}_{i,\mathcal{J}}=F_{d_{k},\mathcal{J}}$ where $d_{k}=i$. In MHCD, the common and individual sub-files should be updated when some of their bits are transmitted  (Step $(3.c.ii.E)$, $(3.c.iii)$, $(4.b.ii.E)$ in Algorithm 3).

For each group $\mathcal{Q}_{t,\mathcal{T}}$, if $|\{k\in[1:K]:d_{k}\in\mathcal{T}\}|-|\mathcal{T}|\leq t$, it can be seen that 
\begin{align*}
\mathcal{Q}_{t,\mathcal{T}} & =\{\mathcal{S}\subseteq[1:K]:\underset{s\in\mathcal{S}}{\cup}\{d_{s}\}=\mathcal{T},\mathcal{S}\cap\mathcal{L}\neq\emptyset,|\mathcal{S}|=t+1\}\\
 & =\{\mathcal{S}\subseteq[1:K]:\underset{s\in\mathcal{S}}{\cup}\{d_{s}\}=\mathcal{T},|\mathcal{S}|=t+1\}.
\end{align*}
For such group the codes in decMAN and in~\cite{exactrateuncoded} are identical and such that we cannot get multicasting opportunities from the latter one. Hence, we consider the group
 $\mathcal{Q}_{t,\mathcal{T}}$, where $|\{k\in[1:K]:d_{k}\in\mathcal{T}\}|-|\mathcal{T}|>t$. The code for this group is $\underset{s\in\mathcal{S}}{\oplus}\mathcal{Y}_{d_{s},\mathcal{S}\setminus\{s\}}$ for each $\mathcal{S}\in \mathcal{Q}_{t,\mathcal{T}}$. In the following, we will introduce the construction of each transmitted $\mathcal{Y}_{i,\mathcal{J}}$ where $(i,\mathcal{J})\in \{(d_{s},\mathcal{\mathcal{S}}\setminus\{s\}):\mathcal{S}\in\mathcal{Q}_{t,\mathcal{T}},s\in\mathcal{S}\}$.
 
Compute $a_{t,\mathcal{T}}=\min_{\mathcal{S}\in\mathcal{Q}_{t,\mathcal{T}}}\max_{s\in\mathcal{S}}|\mathcal{W}_{d_{s},\mathcal{S}\setminus\{s\}}|$. Then, we focus on each pair $(i,\mathcal{J})$ in 
$$
\{(d_{s},\mathcal{\mathcal{S}}\setminus\{s\}):\mathcal{S}\in\mathcal{Q}_{t,\mathcal{T}},s\in\mathcal{S}\}.
$$ 
Let $\mathcal{U}_{i,\mathcal{J}}=\{k\in[1:K]:d_{k}=i\textrm{ and }k\notin\mathcal{J}\}$ representing the users demanding $F_{i,\mathcal{J}}$. 
If $|\mathcal{W}_{i,\mathcal{J}}|\geq a_{t,\mathcal{T}}$, $\mathcal{Y}_{i,\mathcal{J}}$ is equal to the first $a_{t,\mathcal{T}}$ bits from $\mathcal{W}_{i,\mathcal{J}}$. If $|\mathcal{W}_{i,\mathcal{J}}|< a_{t,\mathcal{T}}$, we borrow $a_{t,\mathcal{T}}-|\mathcal{W}_{i,\mathcal{J}}|$ bits from the higher type sub-files $\mathcal{W}_{i,\mathcal{J}_{1}}$ where $\mathcal{J}_{1}\supseteq\mathcal{J},\mathcal{U}_{i,\mathcal{J}}\cap\mathcal{S}_{1}=\emptyset$. Firstly,  we borrow bits from the sub-files $\mathcal{W}_{i,\mathcal{J}_{1}}$ where $|\mathcal{J}_{1}|=t+1$ and, if this is not enough, bits from higher type will be borrowed. $\mathcal{Y}_{i,\mathcal{J}}$ is formed by all the bits in $\mathcal{W}_{i,\mathcal{J}}$ and the borrowed bits; if $\mathcal{Y}_{i,\mathcal{J}}$ is still shorter than $a_{t,\mathcal{T}}$, we pad $a_{t,\mathcal{T}}-|\mathcal{Y}_{i,\mathcal{J}}|$ zeros at the end of $\mathcal{Y}_{i,\mathcal{J}}$.
 
After coding each group in step $t$, some bits in the common requested sub-files have been transmitted 
but some still remain, which are known by $t$ users and not yet transmitted to the demanders. Hence, we use HCD to transmit all the bits remaining in $\mathcal{D}_{k,\mathcal{J}}$ where $k\in [1:K]$ and $|\mathcal{J}|=t$. In other words, for each $\mathcal{S}\subseteq[1:K]$ of size $|\mathcal{S}|=t+1$ and each $s\in \mathcal{S}$, let $\mathcal{D}'_{s,\mathcal{S}\setminus\{s\}}=\mathcal{D}_{s,\mathcal{S}\setminus\{s\}}$ and if $|\mathcal{D}_{s,\mathcal{S}\backslash\{s\}}|<\max_{s\in\mathcal{S}}|\mathcal{D}_{s,\mathcal{S}\setminus\{s\}}|$, we borrow $\max_{s\in\mathcal{S}}|\mathcal{D}_{s,\mathcal{S}\setminus\{s\}}|-|\mathcal{D}_{s,\mathcal{S}\backslash\{s\}}|$ bits from $\mathcal{D}_{s,\mathcal{J}}$ where $s\notin\mathcal{J}$ and $\mathcal{J}\supseteq\mathcal{S}\setminus\{s\}$. The borrowed bits should be padded at the end of $\mathcal{D}'_{s,\mathcal{S}\setminus\{s\}}$. Similarly, if there are not enough bits to borrow, bit $0$ should be padded at the end of $\mathcal{D}'_{s,\mathcal{S}\setminus\{s\}}$. After the bit borrowing step, we transmit  $\underset{s\in\mathcal{S}}{\oplus}\mathcal{D}'_{s,\mathcal{S}\setminus\{s\}}$ for each $\mathcal{S}\subseteq[1:K]$ of size $|\mathcal{S}|=t+1$.

Note that in MHCD, if we need to borrow $x$ bits from one higher type sub-file, we take its first $x$ bits to let the common requested sub-files as large as possible.

\rule{0.45\textwidth}{0.8pt}

\vspace{-3bp}
\textbf{Algorithm 3} Step $t$ of MHCD
\vspace{-6bp}

\rule{0.45\textwidth}{0.8pt}

\begin{enumerate}
\item \textbf{Input:} $\mathcal{D}_{k,\mathcal{J}}$ for each $k\in[1:K]$
and $\mathcal{J}\subseteq[1:K]$ where $|\mathcal{J}|\geq t$, $\mathcal{W}_{i,\mathcal{J}}$ for
each $i\in\mathcal{N}(\mathbf{d})$ and $\mathcal{J}\subseteq[1:K]$ where $|\mathcal{J}|\geq t$,
$\mathcal{L}$,$\mathbf{d}$, $t_{1}=t_{2}=t+1$;
\item $\mathcal{G}_{t}=\{\mathcal{T}\subseteq\mathcal{N}(\mathbf{d}):1\leq|\mathcal{T}|\leq\min\{t+1,|\mathcal{N}(\mathbf{d})|\}\}.$
\item \textbf{for} each $\mathcal{T}\in\mathcal{G}_{t}$ where $|\{j\in[1:K]:d_{j}\in\mathcal{T}\}|-|\mathcal{T}|>t$,
\begin{enumerate}
\item $\mathcal{Q}_{t,\mathcal{T}}=\{\mathcal{S}\subseteq[1:K]:\underset{s\in\mathcal{S}}{\cup}\{d_{s}\}=\mathcal{T},\mathcal{S}\cap\mathcal{L}\neq\emptyset,|\mathcal{S}|=t+1\};$
\item $a_{t,\mathcal{T}}=\min_{\mathcal{S}\in\mathcal{Q}_{t,\mathcal{T}}}\max_{s\in\mathcal{S}}|\mathcal{W}_{d_{s},\mathcal{S}\setminus\{s\}}|;$
\item \textbf{for} each $(i,\mathcal{J})\in\{(d_{s},\mathcal{\mathcal{S}}\setminus\{s\}):\mathcal{S}\in\mathcal{Q}_{t,\mathcal{T}},s\in\mathcal{S}\}$,
\begin{enumerate}
\item $\mathcal{U}_{i,\mathcal{J}}=\{k\in[1:K]:d_{k}=i\textrm{ and }k\notin\mathcal{J}\};$
\item \textbf{if} $|\mathcal{W}_{i,\mathcal{J}}|\geq a_{t,\mathcal{T}}$, \textbf{then}  $\mathcal{Y}_{i,\mathcal{J}}=$the
first $a_{t,\mathcal{T}}$ bits of $\mathcal{W}_{i,\mathcal{J}}$; \\
\textbf{else, then}
\begin{enumerate}
\item $R_{e}=a_{t,\mathcal{T}}-|\mathcal{W}_{i,\mathcal{J}}|$; $\mathcal{Y}_{i,\mathcal{J}}=\mathcal{W}_{i,\mathcal{J}}$;
\item $\mathcal{B}=\{\mathcal{W}_{i,\mathcal{J}_{1}}:|\mathcal{J}_{1}|=t_{1},\mathcal{J}_{1}\supseteq\mathcal{J},\mathcal{U}_{i,\mathcal{J}}\cap\mathcal{J}_{1}=\emptyset,|\mathcal{W}_{i,\mathcal{J}_{1}}|\neq0\};$
\item $\mathcal{W}_{1}=\textrm{Getbits}(\mathcal{B},R_{e})$;$\mathcal{Y}_{i,\mathcal{J}}=\mathcal{Y}_{i,\mathcal{J}}\cup\mathcal{W}_{1};$
\item $R_{e}=R_{e}-|\mathcal{W}_{1}|$;
\item \textbf{for} each $(i,\mathcal{J}_{1})$ where $\mathcal{W}_{i,\mathcal{J}_{1}}\in\mathcal{B}$, update $\mathcal{W}_{i,\mathcal{J}_{1}}=\mathcal{W}_{i,\mathcal{J}_{1}}\setminus\mathcal{W}_{1}$; \textbf{for} each $(k,\mathcal{J}_{1})$ where $k\in \mathcal{U}_{i,\mathcal{J}}$ and $\mathcal{W}_{i,\mathcal{J}_{1}}\in\mathcal{B}$, update $\mathcal{D}_{i,\mathcal{J}_{1}}=\mathcal{D}_{i,\mathcal{J}_{1}}\setminus\mathcal{W}_{1}$;
\item \textbf{if} $t_{1}<K-1$ and $R_{e}>0$, \textbf{then} $t_{1}=t_{1}+1$ and go to Step (3.c.ii.B);
\end{enumerate}
\item \textbf{for} each $k\in\mathcal{U}_{i,\mathcal{J}}$, update $\mathcal{D}_{k,\mathcal{J}}=\mathcal{D}_{k,\mathcal{J}}\setminus\mathcal{Y}_{i,\mathcal{J}}$;
\end{enumerate}
\item \textbf{for} each $\mathcal{S}\in\mathcal{Q}_{t,\mathcal{T}}$, transmit $\underset{s\in\mathcal{S}}{\oplus}\mathcal{Y}_{d_{s},\mathcal{S}\setminus\{s\}};$
\end{enumerate}
\item \textbf{for} each $S\subseteq[1:K]$ where $|\mathcal{S}|=t+1$,
\begin{enumerate}
\item $b_{\mathcal{S}}=\max_{s\in\mathcal{S}}|\mathcal{D}_{s,\mathcal{S}\setminus\{s\}}|;$
\item \textbf{for} each $s\in\mathcal{S}$,
\begin{enumerate}
\item $\mathcal{D}'_{s,\mathcal{S}\setminus\{s\}}=\mathcal{D}_{s,\mathcal{S}\setminus\{s\}}$;
\item \textbf{if} $|\mathcal{D}_{s,\mathcal{S}\setminus\{s\}}|<b_{\mathcal{S}}$, \textbf{then}
\begin{enumerate}
\item $R_{e}=b_{\mathcal{S}}-|\mathcal{D}_{s,\mathcal{S}\setminus\{s\}}|$; 
\item $\mathcal{B}=\{\mathcal{D}_{s,\mathcal{S}_{1}}:|\mathcal{S}_{1}|=t_{2},\mathcal{S}_{1}\supseteq\mathcal{S}\setminus\{s\},s\notin\mathcal{S}_{1},|\mathcal{D}_{j,\mathcal{S}_{1}}|\neq0\};$
\item $\mathcal{W}_{1}=\textrm{Getbits}(\mathcal{B},R_{e})$; $\mathcal{D}'_{s,\mathcal{S}\setminus\{s\}}=\mathcal{D}'_{s,\mathcal{S}\setminus\{s\}}\cup\mathcal{W}_{1};$
\item $R_{e}=R_{e}-|\mathcal{W}_{1}|$;
\item \textbf{for} each $(k,\mathcal{S}_{1})$ where $\mathcal{D}_{k,\mathcal{S}_{1}}\in\mathcal{B}$, update $\mathcal{W}_{d_{k},\mathcal{S}_{1}}=\mathcal{W}_{d_{k},\mathcal{S}_{1}}\setminus\mathcal{W}_{1}$ and $\mathcal{D}_{k,\mathcal{S}_{1}}=\mathcal{D}_{k,\mathcal{S}_{1}}\setminus\mathcal{W}_{1}$;
\item \textbf{if} $t_{2}<K-1$ and $|R_{e}|>0$, \textbf{then} $t_{2}=t_{2}+1$ and go to Step (4.b.ii.B);
\end{enumerate}
\end{enumerate}
\item transmit $\underset{s\in\mathcal{S}}{\oplus}\mathcal{D}'_{s,\mathcal{S}\setminus\{s\}};$
\end{enumerate}
\end{enumerate}

\rule{0.45\textwidth}{0.8pt}

\rule{0.45\textwidth}{0.8pt}

\vspace{-3bp}
\textbf{Algorithm 4} $\textrm{Getbits}(\mathcal{B},R_{e})$
\vspace{-6bp}

\rule{0.45\textwidth}{0.8pt}
\begin{enumerate}
\item $\mathcal{C}=\emptyset$; $a=1$; $b=1$;
\item \textbf{if} $\sum_{e\in\mathcal{B}}|e|\leq R_{e}$, \textbf{then} $\mathcal{C}=$all the bits of all the sub-files in $\mathcal{B}$; \\
\textbf{else, then}
\begin{enumerate}
\item Sort the sub-files in $\mathcal{B}$ by length where $\mathcal{B}(1)$ represents the sub-file with the max length while $\mathcal{B}(|\mathcal{B}|)$ represents the one with min length.
\item \textbf{if} $b\leq |\mathcal{B}(a)|$, \textbf{then } $\mathcal{C}=\mathcal{C}\cup\{\textrm{the }b^{\textrm{th}}\textrm{ bit of }\mathcal{B}(a)\}$; 
\item \textbf{if} $|\mathcal{C}|=R_{e}$, \textbf{then }Output $\mathcal{C}$; \\
\textbf{else if} $a=|\mathcal{B}|$, \textbf{then }$a=1$, $b=b+1$ and go to Step (2); \\
\textbf{else, then} $a=a+1$ and go to Step (2);
\end{enumerate}
\end{enumerate}

\rule{0.45\textwidth}{0.8pt}

\begin{figure}
\centering{}
\includegraphics[scale=0.5]{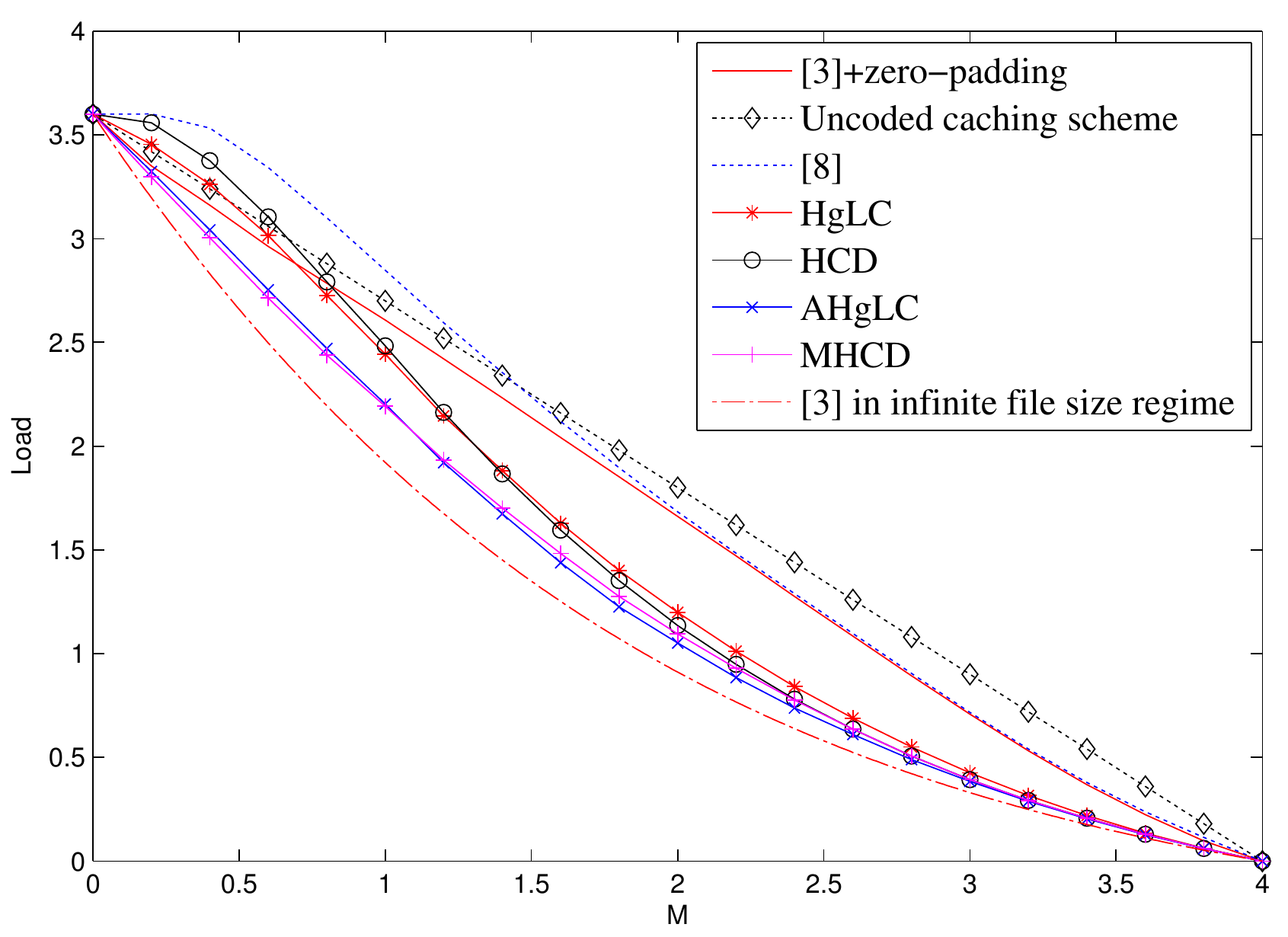}
\caption{Average memory-load (uniform demand) for a decentralized system with $N=4$,
$K=8$ and $F=400$.}
\label{fig:excen}
\end{figure}

\section{Numerical Results and Complexity Analysis}
\label{sec:num}
\paragraph*{Numerical Results} We consider a decentralized caching system with $N=4$, $K=8$ and $F=400$. We assume that the demand distribution is uniform. We compare the average load-memory tradeoffs of the two proposed schemes to the conventional uncoded caching scheme, the caching scheme in~\cite{exactrateuncoded}, HCD and HgLC implemented in the finite file size regime and the scheme in~\cite{finiteanalysis}.  In the placement of uncoded caching schemes, each user stores the same $MF/N$ bits of each file and in the delivery phase the server transmits the remaining part of each requested file. For the other schemes, in the placement phase, each user randomly, uniformly and independently stores $MF/N$ bits of each file. There are some trivial modifications in the delivery phase. First of all, for HCD and the schemes in~\cite{finiteanalysis},  in the simulation we directly transmit the non-cached bits of each demanded file. Secondly, we use Algorithm~2 instead of MDS parity-check matrix for HgLC. Since the file size adapted to the scheme in~\cite{newfinitefilesize} is constrainted, we do not draw the tradeoff of this scheme. More precisely, $F/\binom{K_{1}}{K_{1}M/N}$ should be an integer where $K_{1}$ is a parameter which can be chosen in $[2,\infty)$. Hence, for $M=1$, the possible file sizes are $4$, $28$, $220$, $1820$ and so on.  In addition, we also draw the optimal memory-load tradeoff  in infinite file size regime with uniform demand proposed in~\cite{exactrateuncoded} (given in~\eqref{eq:exactrateuncoded result}) as an outer bound. For each tradeoff point, we use Monte Carlo experiments to simulate $5000$ realizations of the placement. For each placement, we randomly generate one demand vector. In Fig.~1, we can see the two proposed schemes outperform the existing ones. Furthermore, when $M$ is less than $1$ MHCD outperforms AHgLC while when $M$ is larger than $1$, AHgLC performs better. In addition, we also compare our schemes to~\cite{newfinitefilesize} with $F=220$, $K=8$, $N=4$, $M=1$. The average load given by~\cite{newfinitefilesize} is $3.49$ while the loads of AHgLC and MHCD  are $2.2072$ and $2.27337$, respectively.

One should note that when $F$ tends to infinity, MHCD can reach the outer bound because when $F$ is infinity, sub-files demanded by the same number of users have the same length such that the borrowing bit step is not needed in MHCD. Hence, it is equivalent to the scheme in~\cite{exactrateuncoded}.

\paragraph*{Complexity Analysis} It can be seen that HCD and MHCD needs at most $O(2^{2K})$ operations while the complexities of HgLC and AHgLC are both $\mathcal{O}(K|\mathcal{V}|^{2})\equiv \mathcal{O}(K[KF(1-{M}/{N})]^{2})\leq \mathcal{O}(K^{3}F^{2})$ where $|\mathcal{V}|$ represents the total number of nodes in the graph. Hence, when $K$ is large and $F$ is not very large, the complexities of HgLC and AHgLC are lower than HCD and MHCD. When $F$ is large and $K$ is not very large, the complexities of HCD and MHCD are lower than HgLC and AHgLC.

\section{Conclusion and Further Work}
\label{sec:conclusions}
We investigated the decentralized caching
problem with finite file size 
and proposed two novel delivery methods leveraging multicasting opportunities. Numerical results showed that in the uniform demand case our proposed schemes outperforms previous schemes in terms of the average memory-load tradeoff.
Further work includes testing 
the two proposed delivery schemes, with suitable random placement strategies, 
with different demand distributions.
\paragraph*{Acknowledgments}
The work of K. Wan and D. Tuninetti is supported by Labex DigiCosme and in part by NSF~1527059, respectively.



\bibliographystyle{IEEEtran}
\bibliography{IEEEabrv,IEEEexample}

\end{document}